\documentclass[useAMS,usenatbib]{mn2e}
\usepackage{amssymb}
\usepackage{amsmath}
\usepackage[dvips]{graphicx}
\usepackage{epsfig}
\usepackage{multicol}

\date{Accepted ... Received ...; in original form ...}
\title{No visible optical variability from a relativistic blast wave encountering a wind-termination shock}
\author[Van Eerten et al.]{H.J. van Eerten$^{1}$\thanks{E-mail: H.J.vanEerten@uva.nl}, Z. Meliani$^2$, R.A.M.J. Wijers$^1$, R. Keppens$^{2,3,4}$\\
$^{1}$Astronomical Institute 'Anton Pannekoek', PO box 94248, 1090 SJ Amsterdam, the Netherlands\\
$^{2}$Centre for Plasma Astrophysics, K.U. Leuven, Celestijnenlaan 200B, 3001 Leuven, Belgium\\
$^{3}$FOM-Institute for Plasma Physics Rijnhuizen, Nieuwegein, The Netherlands\\
$^{4}$Astronomical Institute, Utrecht University, The Netherlands}
\begin{document}
\maketitle
\begin{abstract}
Gamma-ray burst afterglow flares and rebrightenings of the optical and X-ray light curve have been attributed to both late time inner engine activity and density changes in the medium surrounding the burster. To test the latter, we study the encounter between the relativistic blast wave from a gamma-ray burster and a stellar wind termination shock. The blast wave is simulated using a high performance adaptive mesh relativistic hydrodynamics code, \textsc{amrvac}, and the synchrotron emission is analyzed in detail with a separate radiation code. We find no bump in the resulting light curve, not even for very high density jumps. Furthermore, by analyzing the contributions from the different shock wave regions we are able to establish that it is essential to resolve the blast wave structure in order to make qualitatively correct predictions on the observed output and that the contribution from the reverse shock region will not stand out, even when the magnetic field is increased in this region by repeated shocks. This study resolves a controversy in recent literature.
\end{abstract}
\section{Introduction}
Gamma-ray Burst (GRB) afterglows are produced when a relativistic blast wave interacts with the circumstellar medium around the burster and emits nonthermal radiation. (for reviews, see \citealt{Piran2005, Meszaros2006}) The general shape of the resulting spectra and light curves can be described by combining the self-similar Blandford-McKee (BM) model \citep{Blandford1976} for a relativistic explosion with synchrotron radiation emission from a relativistic electron population accelerated into a power law distribution at the shock front. This model describes a smooth synchrotron light curve, with the slope of the curve a function of the power law slope of the accelerated electrons and of the density structure of the surrounding medium \citep{Meszaros1997, Wijers1997}.

This picture however, is far from complete and with the increasing quality of the available data (e.g. from \emph{swift}) more deviations from the standard of a smoothly decaying (in the optical and X-ray) light curve are being found, for example in the shape of flares \citep{Burrows2005, Nousek2006, OBrien2006} in the X-ray afterglows and early optical variability \citep{Stanek2006}.

Along with prolonged inner engine activity, changes in the surrounding density structure have often been suggested as a cause of this variability \citep{Wang2000, Lazzati2002, Nakar2003}. The details of the shape of the surrounding medium have therefore been the subject of various studies (e.g. \citealt{vanMarle2006}), as well as the hydrodynamics of a relativistic blast wave interacting with a complex density environment \citep{Meliani2007b}. Two recent studies combine a description for the structure of the blast wave after encountering a sudden change in density, like the wind termination shock of a Wolf-Rayet star, with an analysis of the emitted synchrotron radiation that is a result of this encounter (\citealt{Nakar2007}, hereafter NG and \citealt{Peer2006}, hereafter PW), but arrive at different conclusions. A short transitory feature in the observed light curves (at various wavelengths) is predicted by PW, whereas NG conclude that any sudden density change of arbitrary size will result in a smooth transition. The purpose of this paper is to resolve this discrepancy in the literature by performing, for the first time, a detailed analysis of the radiation produced by a blast wave simulated with a high performance adaptive-mesh refinement code. For this analysis, we use the radiation code described in \citet{vanEerten2009} and the \textsc{amrvac} relativistic magnetohydrodynamics (RHD) code (\citealt{Keppens2003, Meliani2007}). We take special care to perform our simulation at a sufficiently high spatial and temporal resolution, such that a transitory feature, if any, is properly resolved.

In section \ref{initial_section} we will first describe the setup and technical details of our simulation run. In section \ref{results_section} we will discuss the resulting optical light curve and the fluid profile during the encounter. Our numerical results confirm those of NG. However, by following the same approximations for the shock wave dynamics as PW, who approximate the different shocked regions by homogeneous slabs, we find that we are able to reproduce their result of a rebrightening of the afterglow curve. In section \ref{slabs_section} we argue how this illustrates the importance of resolving the downstream density structure. After that we separately discuss in section \ref{RS_section} the contribution of the reverse shock that is triggered when the blast wave hits a density discontinuity, as it is the main transitory phenomenon during the encounter. This contribution is overestimated by PW and assumed similar in behavior to that of the forward shock in NG. Since both NG and PW do not invoke electron cooling in their arguments and optical flashes, if any, occur at observer frequencies that are orders of magnitude below the cooling break, we will not enable electron cooling in our radiation code. We summarize our results in section \ref{summary_section}.

\section{Initial conditions}
\label{initial_section}

We will study the case of a massive ($M \gtrsim 25 M_\odot$), low metallicity ($Z \backsim 0.01 Z_\odot$) progenitor star. During its Wolf-Rayet phase (lasting $\backsim 10^6$ years) a stellar wind is produced, which determines the shape of the circumstellar medium. The typical mass-loss rate is approximately $\dot{M} \backsim 10^{-6} M_\odot \textrm{ yr}^{-1}$ and the typical wind velocity $v_w \backsim 1000 \textrm{ km s}^{-1}$. Because the stellar wind flow is supersonic, a shock is produced. A simple schematic description of the circumstellar medium (where we ignore complications such as the influence of photo-ionization) consists therefore of (starting near the star and moving outwards) a free-flowing stellar wind region, a density jump separating the stellar wind region from a homogenized region influenced by the reverse shock, a contact discontinuity followed by a region shocked by the forward shock. The forward shock front then separates the shocked medium from the unshocked interstellar medium (ISM). 

Following the GRB explosion, a relativistic blast wave is sent into this environment. For the typical progenitor values above, an ISM number density $n_{ISM} \backsim 10^3 \textrm{ cm}^{-3}$ and a GRB explosion energy of $E = 10^{53}$ erg, this blast wave will only encounter the first discontinuity during its relativistic stage. The discontinuity will be positioned at $R_0 = 1.6 \cdot 10^{18}$ cm and corresponds to a jump in density of a factor 4. Before the jump the radial density profile is given by $n(r) = 3 \cdot ( r / 1\cdot10^{17} )^{-2} \textrm{ cm}^{-1}$, and after the jump by the constant $n(r) = 4 \times 3 \cdot (R_0 / 1 \cdot 10^{17} )^{-2} \textrm{ cm}^{-1}$. These exact values are chosen to conform to PW.

We have run a number of simulations of relativistic blast waves hitting the wind termination shock at $R_0$. The initial fluid profile is generated from the impulsive energy injection BM solution with the parameters described above for the explosion energy and circumburst density, keeping the adiabatic index fixed at 4/3. The starting time is taken when the shock Lorentz factor is 23. The blast wave will hit the discontinuity when its Lorentz factor is $\backsim 22.27$, at an explosion lab frame time $t_{enc} = 5.34 \cdot 10^7$ seconds (with $t = 0$ set to the start of the explosion). This time corresponds to $\backsim 0.3 $ days for radiation coming from the shock front in observer time (which is taken to be zero at the start of the explosion). To completely simulate the encounter, we will follow the evolution of the blast wave from $5 \cdot 10^6$ seconds to $6.4 \cdot 10^7$ seconds and will store enough output to obtain a temporal resolution (in lab frame simulation time) $d t$ of $1.56 \cdot 10^3$ seconds.

For the outer boundary of the computational grid we take $6 \cdot 10^{18}$ cm, enough to completely capture the shock profile during the encounter even if it were to continue at the speed of light. In order to resolve the shock wave, even at its smallest width at Lorentz factor 23, we take 10 base level cells and allow the adaptive mesh refinement routine to locally double the resolution (where needed) up to 17 times. This implies an effective resolution $dr \backsim 6.3 \cdot 10^{11}$ cm and effectively 1,310,720 grid cells. 

Three simulations were performed using the initial conditions from PW (along with some at lower resolutions, to check for convergence): a test run with stellar wind profile only (and no discontinuity), one with a density jump of 4 and one with a far stronger density jump of 100. Although density jumps much larger than 4 may be feasible (see \citealt{vanMarle2006}, for an example scenario where the progenitor star has a strong proper motion -the relativistic blast wave will then be emitted into a stellar environment that takes the shape of a bow shock), this is not the main motivation for the factor 100 simulation run. The primary focus is on establishing if the lack of an observer effect in the light curve persists for general values of the density jump.

To study relativistic as well as ultra-relativistic blast waves, in addition to the Lorentz factor 23 scenario we have also performed two simulations (one with jump and a test run without) where we moved the density jump outward to $3\cdot10^{19}$ cm, while keeping the other parameters equal. In this scenario the blast wave encounters the jump when it has a shock Lorentz factor $\backsim 5$.

The simulation output is then analyzed using the radiation code for an observer at a distance of $1 \cdot 10^{28}$ cm. The microphysics of the shock acceleration is captured by a number of ignorance parameters. The fraction of thermal energy residing in the small scale downstream magnetic field is $\epsilon_B = 0.01$, the fraction of thermal energy in the accelerated particle distribution $\epsilon_E = 0.1$, the number of power law accelerated electrons as a fraction of the electron number density $\xi_N = 1$ and the slope of this power law $p = 2.5$. Again these values are chosen to match PW.

\section{Light curve and shock profile}
\label{results_section}

The discussion below refers to the shock Lorentz factor 23 scenario. The Lorentz factor 5 simulations lead to qualitatively similar light curves and will therefore not be discussed in further detail. The transition then takes extremely long due to the longer dominance of earlier emission. These simulations confirm that the results presented hold for relativistic blast waves as well, not just for ultra-relativistic blastwaves.

\begin{figure}
%\includegraphics[angle=0, width=0.32\textwidth]{spectrum1.eps}
%\hspace{10mm}
\includegraphics[width=0.49\textwidth]{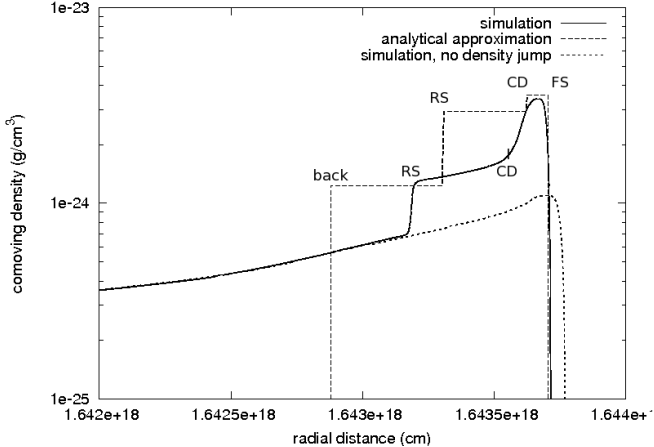}
\caption{A snapshot of the comoving density profile at 17 refinement levels of the fluid at emission time $t_e = 5.48578 \cdot 10^7$ s, for the factor 4 increase in density. The different regions are clearly visible. From left to right we have: up to the steep rise the region not yet influenced by the encounter, the plateau resulting from the passage of the reverse shock, and starting at the gradual rise the region of the forward shock. The front part of the forward shock region is again homogeneous in density, showing the difference between the idealized BM solution and actual simulation results. The flat part of the forward shock region (smallest, rightmost region) is resolved by $\backsim 100$ cells.}
\label{snapshot_figure}
\end{figure}

Directly after hitting the discontinuity, the blast wave splits into three regions. The innermost region, up to the reverse shock (RS) front remains unaware of the collision. Beyond the RS the plasma gets homogenized up to the contact discontinuity (CD). The region following the contact discontinuity, up to the forward shock (FS) is not homogeneous but will gradually evolve into a BM profile again for a modified value of the circumburst density structure. A snapshot of the shock structure during the encounter is shown in figure \ref{snapshot_figure}. We show comoving density (as opposed to the lab frame density) because the differences between the different regions then stand out more clearly.

\begin{figure}
\includegraphics[width=0.49\textwidth]{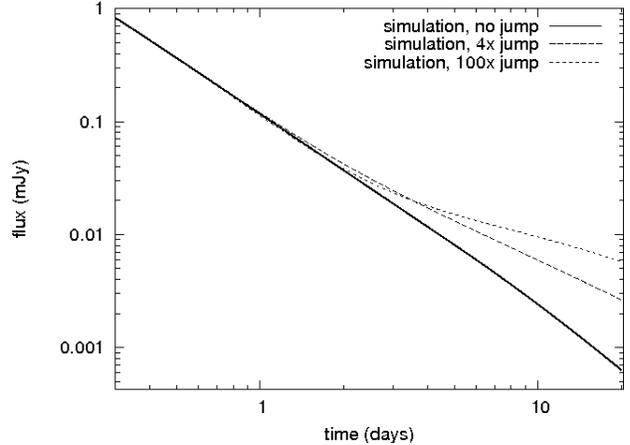}
\caption{The figure shows the resulting optical light curves at $5 \cdot 10^{14}$ Hz, for the cases of a continuous stellar wind environment, a jump of a factor 4 followed by a homogeneous environment and a jump of a factor 100. 50 data points have been devoted to 0.3 - 1 day and 50 data points to the following 19 days. A smooth transition towards the power law behaviour corresponding to a BM shock wave expanding into a homogeneous environment is visible, even for the extreme change in density.}
\label{termination_shock_figure}
\end{figure}

The optical light curves calculated from the simulations are observed at $\nu = 5 \cdot 10^{14}$ Hz, which lies between the synchrotron peak frequency $\nu_m$ and the cooling break frequency $\nu_c$ (it may be helpful to emphasize that here, contrary to shock interaction during the prompt emission phase, $\nu_m$ is found at a similar frequency for both the forward and reverse shock contributions). Because the observer frequency lies well below the cooling break, we ignore the effect of electron cooling. The light curves for the factor 4 and factor 100 density jumps are found in figure \ref{termination_shock_figure}. For complete coverage at the observed times and clarity of presentation, analytically calculated emission from a BM profile with Lorentz factors $>23$ (or $>5$) has been added to that calculated from the simulations. From the light curves we draw the following conclusion: \emph{an encounter between the relativistic blast wave and a wind termination shock does not lead to a bump in the light curve, but instead to a smooth change in slope.} The new slope eventually matches that of a BM solution for the density structure found beyond the discontinuity.

\section{Resolved blast wave versus homogeneous slab}
\label{slabs_section}
\begin{figure}
%\includegraphics[angle=0, width=0.32\textwidth]{spectrum1.eps}
%\hspace{10mm}
\includegraphics[angle=0, width=0.49\textwidth]{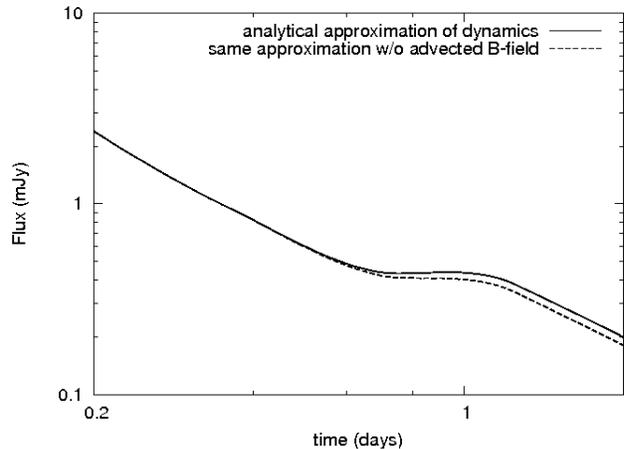}
\caption{Resulting light curves at $5 \cdot 10^{14}$ Hz when our radiation code is applied to the homogeneous slabs approximation of PW, instead of a hydrodynamical simulation. The bottom curve shows the resulting light curve if the magnetic field in the reverse shock region does not contain the additional increase in magnetic field in the reverse shock region. Contrary to the light curves shown in figure \ref{termination_shock_figure}, in \emph{both} cases a clear rise in intensity with respect to the previous level is seen over the course of a few hours, as predicted by PW for homogeneous slabs.}
\label{Peer_lightcurves_figure}
\end{figure}

The optical light curves presented in the above section differ distinctly from those presented by PW in that they show no bumps. This difference in results has to be caused by one or more differences in our assumptions, which are:
\begin{itemize}
 \item PW include both electron cooling and synchrotron self-absorption, while in this paper we have included neither.
 \item We take the magnetic field to be a fixed fraction $\epsilon_B$ of the local thermal energy in all parts of the fluid, even those shocked twice, whereas PW have a magnetic field in the reverse shock region that is slightly higher. This is because they take into account that the dominant magnetic field in the reverse shock region is actually the field advected with the flow from the region shocked once. The newly created field is approximately a factor 1.2 smaller.
 \item We resolve the downstream fluid profile, while PW approximate the different regions behind the shock front by homogeneous slabs of varying density, thermal energy and Lorentz factor. Also, they freeze the fluid Lorentz factors during the encounter.
\end{itemize} 
Since the optical light curve corresponds to an observer frequency sufficiently above the self-absorption critical frequency and sufficiently below the cooling break frequency, neither cooling nor absorption should have any visible effect on the shape of the curve. The fact that cooling is not required for the bump found by PW is also immediately obvious from figure \ref{Peer_lightcurves_figure}, where we have applied our radiation code directly to the homogeneous slabs approximation of PW, with electron cooling disabled. The light curves thus generated \emph{do} show a bump feature after the onset of the encounter (this also provides a check on the internal consistency of both models). To explicitly check the effect of the stronger magnetic field in the reverse shock region we have generated two light curves: one where all the fluid quantities are exactly similar to those of PW and one where we ignored the stronger field in the reverse shock region but kept the field at fixed fraction of the thermal energy (which is the same as that in the forward region in the homogeneous slab approximations, due to pressure balance across the contact discontinuity). As can be seen from the figure, the temporary rise occurs in both cases, with only a marginal difference between the two curves. 

%That the advected magnetic field is not sufficient to explain the rise in flux of a factor $\ge 2$ inferred by PW, can also be seen by comparing the flux for the two cases. From standard scaling arguments one can establish that the flux $F_s$ for the strong field $B_s$ then compares to the flux $F_w$ of the weaker field $B_w$ according to $ F_s / F_w = ( B_s / B_w )^p$. Since the difference between the two fields is a factor 1.2, the resulting difference in flux will be (for $p=2.5$) approximately a factor 1.6. This difference will be smeared out in the resulting light curves so the actual difference between the curves will be less and in any case not sufficient to explain the factor $\ge 2$. 

This brings us to the third difference listed. We conclude that \emph{to determine the visible response of a blast wave to density perturbations, it is crucial to take the radial structure of the blast wave into account}. This (along with establishing the lack of a transitory feature itself) is the main conclusion from this paper and forms an important justification for the kind of detailed approach that we have employed, where the dynamics of the blast wave are simulated using a high performance RHD code, together with a radiation code that accurately probes all local contributions to the synchrotron spectrum. It is also important to emphasize that the bump found by PW is \emph{not} the result of inaccurately modeling the different arrival times for photons arriving from different angles relative to the line of sight, as has been stated in NG. This also can be seen from figure \ref{Peer_lightcurves_figure} which confirms that, for homogeneous slabs, the light curves published by PW are calculated correctly.

The importance of the downstream shock structure can be understood as follows. By taking a homogeneous slab one not only locally overestimates the downstream density, but also the Lorentz factor and thermal energy (and hence the magnetic field). Also, the width of the homogeneous slab is determined by comparison with the downstream density structure \emph{or} the energy density structure \emph{or} the velocity structure, and matching the width to one of these comes at the expense of a lack of similarity to the others. (And finally, keeping the Lorentz factors fixed during the encounter also contributes to the overestimation of the flux emitted during the encounter). Essentially, all this indicates a lack of resolution. The homogeneous slab implies a spatial resolution\footnote{Even though PW identify three different regions during the encounter, this in itself does not imply an improved spatial resolution, since the fluid conditions in each region are connected to each other (and the upstream medium) via shock-jump conditions that strictly speaking require all regions to be directly adjacent at the same position. The simulation snapshot in fig. \ref{snapshot_figure}, shows that the assumption of the reverse shock region being thermalized and isotropic is not unreasonable, but also shows a clear density gradient within the forward shock region.} $\Delta r \backsim R / \Gamma^2$ cm  (with $R$ the blast wave radius and $\Gamma$ the blast wave Lorentz factor), and is therefore in principle only applicable to describe behavior on time scales $\Delta t > \Delta r / c$. This is true in general, not just for simulations, and in our case yields $\Delta t \backsim 1.5$ days at the time of the onset of the encounter. The reason that the homogeneous slab \emph{does} work to describe the general shape of the light curve from the BM blast wave, as was done by \citet{Waxman1997} among others, is that in these cases the slab is used to describe behavior on time scales $\Delta t >> \Delta r / c$ (actually $\Delta t$ arbitrary large, for understanding of asymptotic behavior). But one should for example not expect the homogeneous slab approximation to get the absolute scale right, and indeed it is off by a factor of a few (justifying more detailed calculations like \citealt{Granot2002, vanEerten2009}).
\begin{figure}
%\includegraphics[angle=0, width=0.32\textwidth]{spectrum1.eps}
%\hspace{10mm}
\includegraphics[angle=0, width=0.49\textwidth]{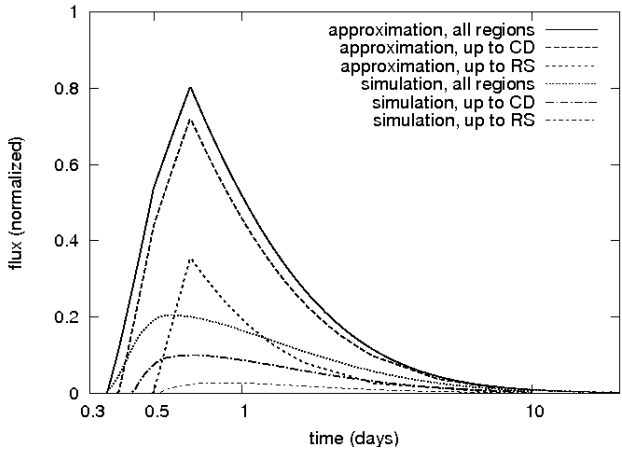}
\caption{Received flux at observer frequency $5 \cdot 10^{14}$ Hz, calculated for a single emission time $t_e = 5.48578 \cdot 10^7$ s (the same time as in fig. \ref{snapshot_figure}). Curves are shown both for the homogeneous slab approximation and for the numerical simulation fluid profile. In each case the contribution from the different regions has been marked: the top curve shows the total flux, the curve below the flux when the contribution from the forward shock region is omitted and for the lowest curve the reverse shock region has been omitted as well. The flux level for the homogeneous slab approximation is much higher than that from the simulation, with (for this particular emission time) the contribution from the reverse shock dominating the total output. At the same emission time, the reverse shock region contribution for the simulation is still significant, but no longer dominant. For the simulation snapshot we have estimated the position of the contact discontinuity, and therefore the edge of the reverse shock region, at the right edge of the plateau, before the onset of the rise in density. (see fig. \ref{snapshot_figure}.)}
\label{single_flash_figure}
\end{figure}

\section{The reverse shock contribution}
\label{RS_section}

In the previous we have established that the reverse shock caused by the encounter with the density perturbation does not cause a rise in the observed light curve. Since this reverse shock has been evoked to explain rebrightening (e.g. by \citealt{Ramirez2005}), it is of interest to look at its contribution in some more detail. In fig. \ref{single_flash_figure}. this contribution (in the optical) is compared directly to the total flux emitted from the shock profile, both for the simulation and for the PW approximation. The important difference is the relative overestimation of the reverse shock region in the PW approximation. The relative contributions for the different regions within either the homogeneous slab or the resolved blast wave simulation of course depend on their relative sizes and therefore on the emission time. Another feature of note is that the homogeneous slabs approximation results in an emission profile that is sharply peaked, whereas the more accurate profile displays a flatter tail and a smoother transition between rise and decay.

The shock structure is also calculated and implemented in NG, starting from the shock jump conditions and assuming homogeneous slabs for the forward and reverse shock region, yet they do not find a temporary rebrightening. This is a consequence of the fact that they set the reverse shock contribution at a fixed fraction of the forward shock contribution, while allowing this forward shock contribution to evolve according to the appropriate BM profile following the density change, as opposed to freezing the shock Lorentz factors during the encounter. That the forward shock determines the shape of the light curve is then imposed as a feature of their model (i.e. in their equation 20) and yields an adequate heuristic description of the light curve found as a result of their simulations.

The difference between the simulations by NG and ours is merely a technical one: instead of a Eulerian code (that can also be used for simulations in more than one dimension, which we will perform in future work), they use a Lagrangian code for the dynamics. The reconstruction of the light curves from the code is equivalent. They also, like us, do not take a slight increase in the magnetic field in the reverse shock region into account. NG provide no information on the spatial and temporal resolution of their simulations.

\section{Summary and conclusions}
\label{summary_section}

We have performed high resolution hydrodynamical simulations of a relativistic blast wave encountering a wind termination shock and have calculated the resulting light curve using the radiation code described in \citet{vanEerten2009}. As a result we have found \emph{no} variability in the optical, not even for very large density changes, for blast waves in the self-similar phase. This renders it very unlikely that observed optical variability in GRB afterglow light curves can be explained from density perturbations in the external medium surrounding the burster, as suggested by e.g. \citet{Wang2000, Lazzati2002, Nakar2003}, PW. This research, however, has been limited to spherically symmetric density perturbations. A second caveat is the assumption of self-similarity for the blast wave approaching the wind termination shock. As demonstrated by \citet{Meliani2007b}, for a termination shock close to the star ($R \backsim 10^{16}$ cm in their simulation, for a short Wolf-Rayet phase), the blast wave structure may still somewhat retain the initial structure of the ejecta (in their simulation, a uniform static and hot shell, i.e. fireball), which may have observable consequences. The latter is however not likely, given the already reasonably strong resemblance between their simulation output during the encounter and ours, where the same shock regions can be identified in the fluid profile with similar values for the physical quantities of interest. Also, if the pre-encounter shock wave is sufficiently different from the self-similar solution this will also have consequences for the global shape and temporal evolution of the observable light curve, and the slope will become markedly different from the one predicted from the BM solution. 

Of the two main explanations for (sometimes quite strong) late optical variability, refreshed or multiple shocks appear to be a far more realistic option than circumburst medium interactions. We are currently performing simulations on multiple interacting shocks to test this alternative hypothesis.

We have compared the results of our simulation to the literature and from a comparison to the approximations and assumptions used by PW and NG especially, we conclude that the fact that we resolve the radial blast wave structure explains the discrepancy between our resuls and those of PW. This, in turn, forms an important justification for the kind of detailed approach that we have employed, where the dynamics of the blast wave are simulated using a high perfomance RHD code, together with a radiation code that accurately probes all local contributions to the synchrotron spectrum. We note that, contrary to what is stated by NG, the calculation of angular smearing of the signal in PW (which in turn was based on \citealt{Waxman1997}) is correct.

\section*{Acknowledgements}

This research was supported by NWO Vici grant 639.043.302 (RAMJW) and NOVA project 10.3.2.02 (HJvE). ZM performed computations on the K.U.Leuven High Performance computing cluster VIC, and acknowledges financial support from the FWO, grant G.0277.08, and from the GOA/2009/009. We would like to thank Asaf Pe'er for feedback and discussion.

%\bsp

\end{document}